\documentclass[aps,prl,twocolumn,superscriptaddress,nofootinbib]{revtex4-2}

\usepackage{graphicx}
\usepackage{amsmath,amssymb}
\usepackage{bm}
\graphicspath{{./}{figures/}}
\makeatletter
\def\input@path{{./}{figures/}}
\makeatother

\newcommand{\figorbox}[2]{%
  \IfFileExists{#1}{\includegraphics[width=\linewidth]{#1}}%
  {\fbox{\parbox[c][0.28\textheight][c]{0.92\linewidth}{\centering\itshape
   #2\\[4pt]\normalfont\footnotesize (figure file not found; drop it in to render)}}}}

\begin{document}

\title{Microwave spectroscopic evidence of a moir\'e-commensurate electron
crystal in monolayer graphene}

\author{A.~Solodovnyk}
\affiliation{Physics Department, Villanova University, Villanova, Pennsylvania 19085, USA}
\author{P.~Colgan}
\affiliation{Physics Department, Villanova University, Villanova, Pennsylvania 19085, USA}
\author{K.~Watanabe}
\affiliation{National Institute for Materials Science, Namiki 1-1, Tsukuba, Ibaraki 305-0044, Japan}
\author{T.~Taniguchi}
\affiliation{National Institute for Materials Science, Namiki 1-1, Tsukuba, Ibaraki 305-0044, Japan}
\author{L.~W.~Engel}
\affiliation{National High Magnetic Field Laboratory, Tallahassee, Florida 32310, USA}
\author{S.~Dietrich}
\email{scott.dietrich@villanova.edu}
\affiliation{Physics Department, Villanova University, Villanova, Pennsylvania 19085, USA}

\date{\today}

\begin{abstract}
We probe the conductivity of the lowest Landau level of a moir\'e-aligned monolayer graphene
device by broadband microwave transmission through a coplanar waveguide ($\sim50$~mK, $15.4$--$18$~T),
isolating the longitudinal channel through a low-minus-high frequency band difference $D(\nu)$. We resolve
a series of minima in $D(\nu)$ that drift with field along lines of fixed carrier density rather than
fixed Landau filling, the hallmark of moir\'e-commensurate electron crystals. The strongest minimum is
present at every field and matches no fractional quantum Hall or integer Hofstadter state. It lies at band
filling $s=1/3$, one electron for every three moir\'e cells---the generalized Wigner crystal that DaSilva,
Jung, and MacDonald predict as the most favorable fixed-density ($t=0$) commensuration. The weaker minima
fall near neighboring commensurate fillings.
\end{abstract}

\maketitle

\begin{figure*}
\includegraphics[width=\linewidth]{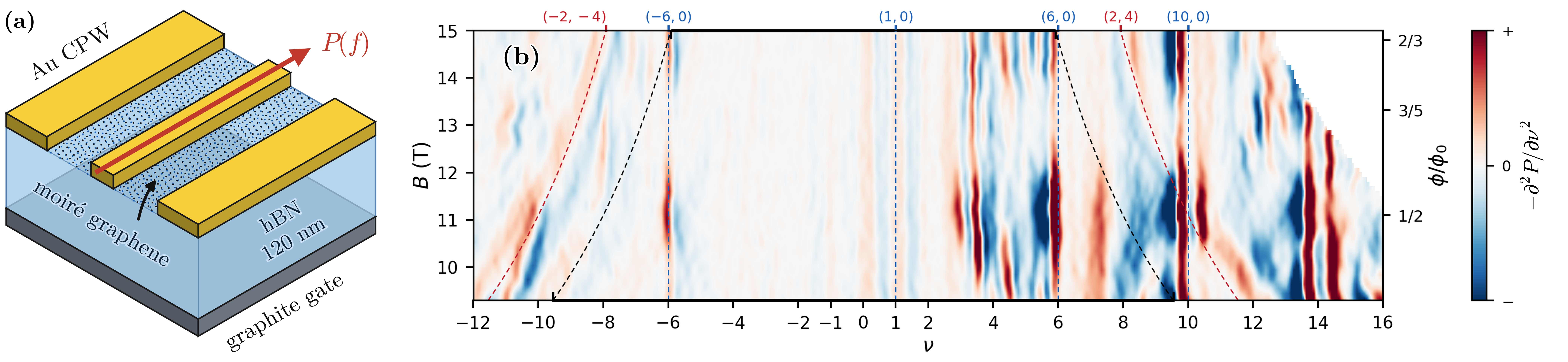}
\caption{\label{fig:fan}\textbf{Device and moir\'e Hofstadter structure.}
(a) Cutaway schematic. Monolayer graphene aligned to the encapsulating hBN (blue hBN, black carbon
atoms) forms a moir\'e superlattice in the slot of a Au CPW, separated from the graphite gate by
$120$~nm of hBN; $S_{21}(f)$ is measured through the CPW center conductor. (b) Second derivative
$-\partial^2 P/\partial\nu^2$ of the fixed-frequency ($3$~GHz) transmission ($P$ in dB) versus filling
$\nu$ and field $B$, following the Diophantine relation, Eq.~\eqref{eq:dioph}. Integer states ($s=0$)
run vertically (blue), sloped gaps ($s\neq0$) fan out (red), and black bars mark the moir\'e band edges
($s=\pm4$). The right axis gives the flux per cell $\phi/\phi_0$, which pinches at $1/2$ ($11.1$~T).}
\end{figure*}

A partially filled Landau level is a setting where an electron liquid and an electron solid have
nearly equal energy. A sufficiently large magnetic field quenches the kinetic energy, so the Coulomb interaction alone
determines which is energetically favorable. In general, a fractional quantum Hall effect (QHE) fluid wins near the
middle of the level, while a Wigner crystal wins toward its edges. Which state appears
at a given filling is a close contest, and small perturbations can decide it. A moir\'e superlattice
is one such perturbation. Its periodic potential lowers the energy of a crystal at first order,
because a solid can lock its density to the superlattice, while a fluid responds only at second
order. A moir\'e potential should therefore tip the balance toward the solid, stabilizing a pinned
electron crystal at fillings where the unmodulated system would form a fluid~\cite{dasilva2016}.

This mechanism selects states commensurate with the superlattice, at simple rational numbers of
electrons per moir\'e cell. In a magnetic field the bookkeeping is fixed by the Diophantine
(Streda--Wannier) relation~\cite{wannier1978,streda1982} for the number of electrons per moir\'e
cell $\nu_P$,
\begin{equation}
\nu_P = t\,\frac{\phi}{\phi_0} + s,
\label{eq:dioph}
\end{equation}
with a Hall (Chern) index $t$, a band-filling (Bloch) index $s$, and the flux per moir\'e cell $\phi$
in units of the flux quantum $\phi_0$. The candidate states sort by the band-filling index
$s$. A state with $s=0$ is a pure Landau-level quantum Hall state, the integer quantum Hall effect when $t$
is an integer and a fractional quantum Hall state, the Laughlin and Jain gaps, when $t$ is fractional; the
half-filled level is instead a compressible composite-fermion sea. A state with integer $s\neq0$ is a
single-particle Hofstadter gap of the fractal spectrum. These integer quantum Hall and Hofstadter gaps
appear in the Wannier diagram measured around half flux, Fig.~\ref{fig:fan}(b).

A fractional $s=p/q$ marks a moir\'e-commensurate state, $p$ carriers per $q$ moir\'e cells. When its Hall index $t$ is fractional
this is the fractal fractional quantum Hall effect, the interaction-driven liquid replicated in the
moir\'e miniband~\cite{wang2015}. When $t$ is an integer it is the moir\'e-commensurate crystal; being
Chern-trivial, its $t$ records only the filled Landau level the carriers are counted from. In the lowest
Landau level the electron crystal $(0,1/3)$ is counted from charge neutrality and the associated hole
crystal $(1,-1/3)$ from the filled level. Identifying the crystal means separating it from its quantum Hall, Hofstadter, and fractal neighbors,
because all of these states can fall at the same filling.

Electron crystals in moir\'e and Landau-quantized systems have been detected chiefly through their
charge, by real-space imaging of generalized Wigner crystals and field-induced Wigner
crystals~\cite{li2021,tsui2024} as well as by quantized transport of moir\'e-driven and quantum anomalous
Hall crystals~\cite{cai2023,park2023,lu2024}. What sets a carrier solid apart, though, is how it
dissipates energy under an applied drive, a dynamical response that charge-based detection does not reach
and that no dynamical probe has yet resolved in these systems. Meanwhile, microwave spectroscopy has long
been the standard probe of pinned electron solids
in conventional GaAs-hosted two-dimensional electron systems, where it resolves their dissipative
response and the pinning-mode resonance it carries~\cite{andrei1988,williams1991,chen2006,chitra2001}, but
it has not been applied to electron crystals in mechanically exfoliated van der Waals devices.

Here we study microwave dissipation in a moir\'e-aligned monolayer graphene device in its lowest
Landau level ($|\nu|\leq1$) at high field and millikelvin temperature. The longitudinal response across the
level is a disorder-broadened dome whose width reflects a density inhomogeneity smaller than the zero-field
transport value. Riding on the dome are intermittent low-dissipation minima that hold a fixed
per-cell density and recur on both carriers at every field, matching no fractional quantum Hall or integer
Hofstadter filling. The strongest is consistent with a $1/3$ crystal of one carrier per three moir\'e
cells, an electron crystal on the electron-doped side and its charge-conjugate hole crystal on the hole
side. The remaining minima point to further commensurate crystals, though we cannot exclude integer
Hofstadter gaps or a mixture of both.

\textit{Device and measurement.---} The device is a monolayer graphene flake encapsulated in hexagonal
boron nitride (hBN), gated by a graphite back gate, and placed in the slot of a Au coplanar
waveguide (CPW), sketched in Fig.~\ref{fig:fan}(a). The graphene is crystallographically aligned to the
hBN ($\theta\approx0^\circ$), forming a triangular moir\'e superlattice of period
$\lambda\approx14.7$~nm and unit-cell area $186.3$~nm$^2$ (S1). At our fields the flux per cell is a
large fraction of a quantum, $\phi/\phi_0=0.69$--$0.81$ across $15.4$--$18$~T, placing the device deep
in the Hofstadter regime~\cite{hofstadter1976}. Figure~\ref{fig:fan}(b) shows the resulting Wannier diagram:
integer-QHE gaps run vertically, the sloped moir\'e Hofstadter gaps fan out from the $s=\pm4$ band
edges, and the filling lines pinch at half flux ($11.1$~T), which fixes $\phi/\phi_0=0.045\,B$. The
lowest Landau level we probe lies far below those band edges. We measure the transmission coefficient
$S_{21}(f)$ over $f=0.5$--$9$~GHz with a vector network analyzer while sweeping the gate voltage, in
linear response. Fabrication and calibration are detailed in S1.

The graphene sits in the CPW gap, so changes in the dissipative conductivity attenuate the
transmission. We quote the transmitted power in decibels, $P\equiv10\log_{10}|S_{21}|^2$, with $S_{21}$
the complex transmission coefficient, and normalize it to the value at charge neutrality,
$P_{\rm n}(f,\nu)=P(f,\nu)-P(f,0)$. At each filling
$P_{\rm n}(f)$ rolls over from a low-frequency plateau to a high-frequency one, and the
normalized map $P_{\rm n}(f,\nu)$ traces this crossover across the lowest Landau level
(S1). To isolate the
longitudinal response versus filling we form the band difference
\begin{equation}
D(\nu) = \langle P_{\rm n}\rangle_{\rm low} - \langle P_{\rm n}\rangle_{\rm high},
\end{equation}
with the low band covering $0.5$--$1$~GHz and the high band covering $7$--$9$~GHz. The transmission is
dominated by the Hall (transverse) admittance, which our earlier work established is
frequency-independent across these bands~\cite{dietrich2025}, so it cancels in the band difference and
leaves only the frequency-dependent longitudinal response $\sigma_{xx}$. $D(\nu)$ is our primary
observable. The filling is set from the gate voltage, with the gate lever arm (carrier density per
unit gate voltage) and charge-neutrality point
calibrated on each scan from the $\nu=\pm1$ states, and the flux per cell fixed by the secondary-Dirac
spacing (S1), consistent with the half-flux pinch above.

\begin{figure}
\includegraphics[width=\linewidth]{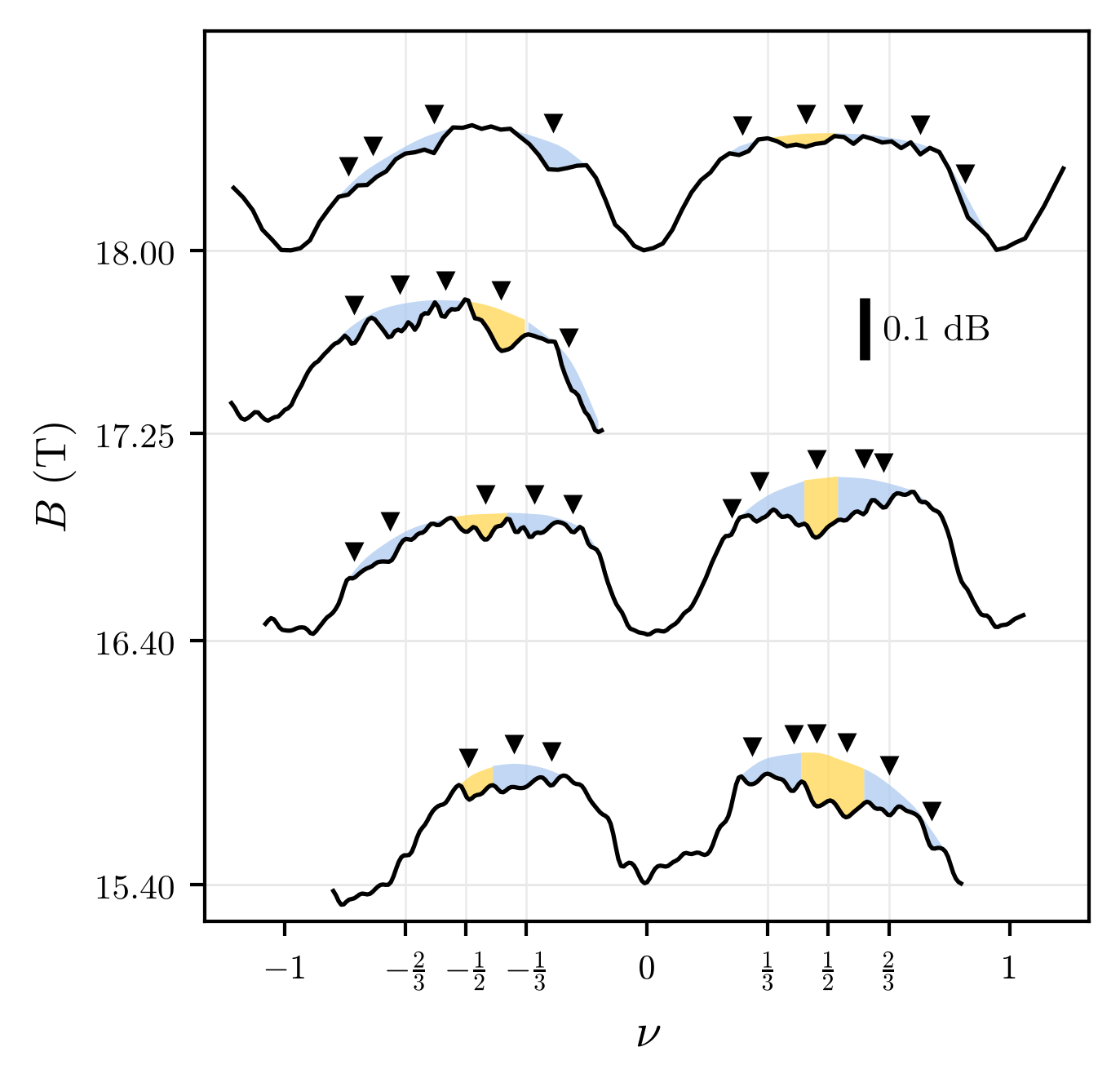}
\caption{\label{fig:dome}\textbf{Band-difference dome.} Band difference $D(\nu)$ across the lowest Landau
level at four fields (bottom to top, $15.40$, $16.40$, $17.25$, $18.00$~T, with traces offset by field),
each a density-inhomogeneity--broadened dome on both sides of charge neutrality. Shaded regions trace
the minima beneath the smooth dome and show how far each falls below it, gold for minima
consistent with the $1/3$ commensuration and light blue for the other minima; triangles mark the minima. The $0.1$~dB bar sets the vertical scale.}
\end{figure}

\textit{Longitudinal response.---} As Fig.~\ref{fig:dome} shows, at each of our four fields $D(\nu)$ forms a broad dome on each side of charge
neutrality across the lowest Landau level, rising from the gapped $\nu=0$ region to a maximum near half
filling and falling toward the $\nu=\pm1$ gaps. Its width is set by the disorder broadening of the Landau level.
Fitting each side with a broadened-level profile of width $\eta$ and converting to a density,
$\delta n=\eta\,n_\phi$ with $n_\phi=eB/h$, gives a field-independent broadening,
$\delta n\approx1.3\times10^{10}$~cm$^{-2}$ across all magnetic fields (the dimensionless width changes
only because $n_\phi\propto B$). This is a fixed density scale, not a magnetic-length scale, and it is
the same charge inhomogeneity that broadens the $B=0$ charge-neutrality point, though smaller than the
full zero-field transport width $\delta n_{\rm tr}\approx3.4\times10^{10}$~cm$^{-2}$~(S2). The dome is therefore produced by the density-inhomogeneity broadening of
the lowest Landau level.

On top of the smooth dome, $D(\nu)$ shows localized
minima where the dissipation is suppressed below the envelope. Figure~\ref{fig:dome} stacks $D(\nu)$ at the
four fields; on each trace a triangle marks an extracted minimum, and the shading measures how far the
minimum falls below the dome, gold where it is consistent with the $1/3$ commensuration and light blue
otherwise. The minima appear at every field, on both the electron and hole sides. The most prominent lies on the electron side at $16.4$~T, at $\nu=+0.468$. Each minimum's filling is taken from $D(\nu)$ itself rather than
assumed at a target fraction, so the fillings are an output of the data. These fillings are not the
expected fractional QHE values. The nearest low-order Jain fractions, $3/7=0.4286$ and $4/9=0.4444$, miss
the most prominent one, and $\nu=1/2$ is excluded on its high side. The minima also do not deepen with field,
as a fractional QHE gap would. Their origin lies beyond the fractional QHE.

\begin{figure}
\includegraphics[width=\linewidth]{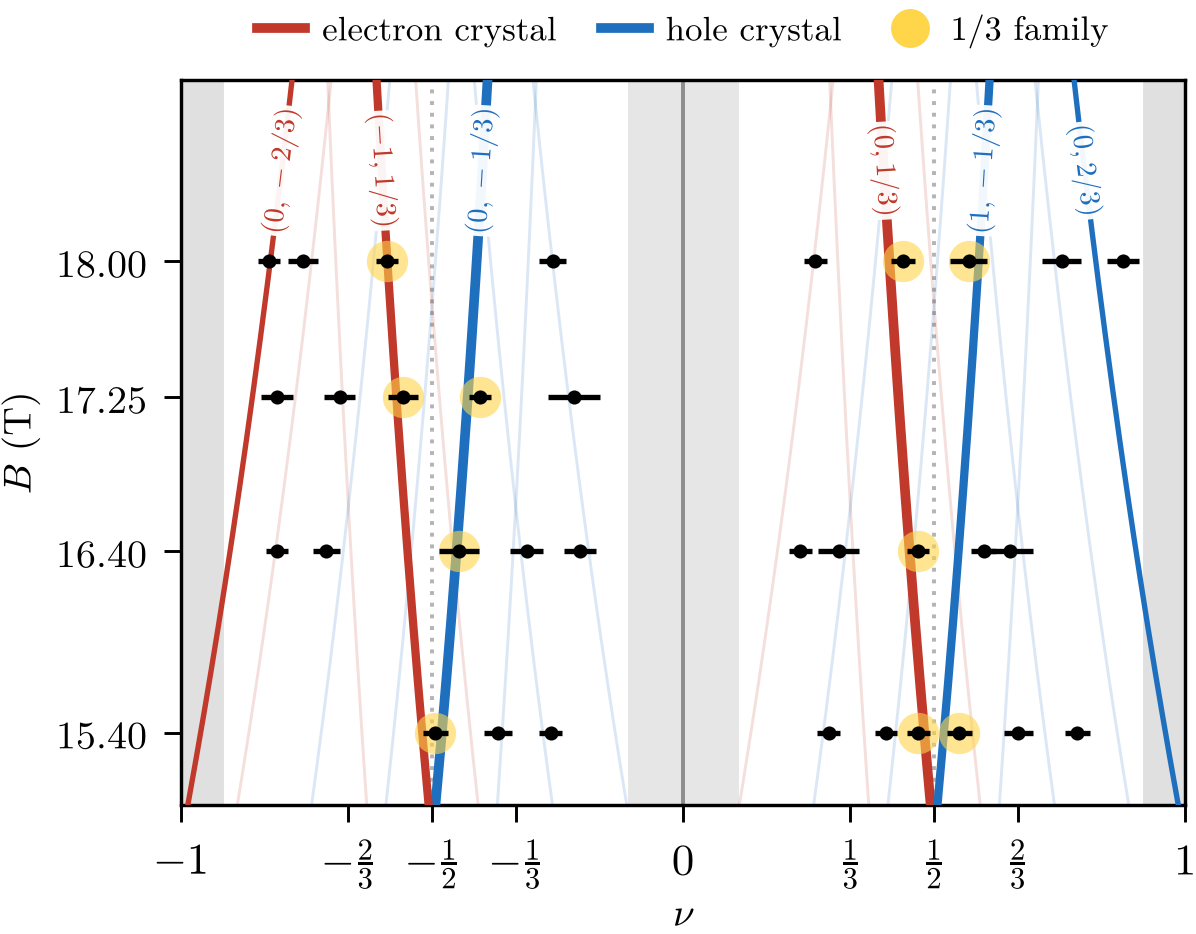}
\caption{\label{fig:crystal}\textbf{Commensurate crystal lines against the observed minima}, on the
signed $\nu$--$B$ plane, each labeled by its Diophantine indices $(t,s)$ and colored by the crystallizing
carrier (red electron, blue hole). Thick lines are the dominant $1/3$ crystals $(0,\pm1/3)$, with thickness
scaling as the expected commensuration strength. The $(1,-1/3)$ and $(-1,1/3)$ lines are their
particle--hole partners, $(0,\pm2/3)$ the honeycomb crystals, and faint lines the weaker fractions
$s=1/4,2/5,1/2,3/5$. Black points are the measured minima, with bars giving the filling uncertainty
and gold halos marking the minima on the $1/3$ family. Dotted verticals mark $\nu=\pm1/2$ and shaded bands
the $\nu=0$ and $\nu=\pm1$ gaps.}
\end{figure}

\textit{Origin of the minima.---} We ask which origin threads the minima across our four
fields with the fewest lines. Each candidate moves through filling as $\nu=\nu_P/(\phi/\phi_0)=t+s/(\phi/\phi_0)$, and the band-filling
index $s$ sets how~(S7). A
Landau-level state ($s=0$), integer QHE or Jain, stays at a fixed $\nu=t$. A moir\'e-commensurate crystal
carries a small fractional $s$, so its filling moves only weakly with field, as $s/(\phi/\phi_0)$,
approaching $\nu=t$ (the level its carriers are counted from) at high field. An integer Hofstadter
gap ($|s|\ge1$) follows the same $1/B$ form but sweeps several times farther across filling over our range. Figure~\ref{fig:crystal} draws the
commensurate-crystal tracks against the observed minima. The $1/3$-family minima (gold) move together
along the fixed-density $(0,\pm1/3)$ lines. They hold a constant per-cell filling $\nu_P=1/3$ and so drift
as $1/B$, the strongest moving from $\nu\approx+0.468$ at $16.4$~T toward charge neutrality at higher field,
and they recur on both carriers.
It is this tracking along a fixed-density line, not any single minimum, that marks the commensuration: the
minima follow a line of constant per-cell filling and move with field as $1/B$, rather than sitting at
a fixed Landau filling, and $s=1/3$ is the commensuration DaSilva, Jung, and MacDonald~\cite{dasilva2016} identify as most
favorable at these fillings~(S3). Whether this is the crystal rather than a competing fluid or gap we
consider next from the physical response.

For that dominant $s=1/3$ feature the fixed-filling and Hofstadter origins are both disfavored. A fractional QHE or
composite-fermion state would sit at a fixed Landau filling, not drift with field as ours does. We see no
fractional-QHE comb elsewhere in the level, and the dissipative rollover frequency stays flat
with no composite-fermion imprint~(S6). The microwave excitation power also favors a robust pinned phase over a
fractional QHE gap, which is small and thermally activated~\cite{boebinger1985} and readily washed out by
the elevated electron temperature of the drive, whereas the moir\'e-commensurate crystal is stabilized to
first order by the superlattice potential. An integer Hofstadter gap is excluded
\emph{for this feature} not by the $\nu\sim1/B$ trend it would share but by its magnitude. The nearest
low-order line $(t,s)=(-1,+1)$ sweeps by $\approx0.21$ over our range, several times the feature's weak
drift along the gentler $(0,1/3)$ track.

A general filling constraint then fixes the character of that state. A gapped state at a fractional
filling per moir\'e cell cannot be a simple insulator: it must either break translational symmetry,
forming a crystal, or carry intrinsic topological order (a Lieb--Schultz--Mattis-type filling
constraint~\cite{lsm1961}). The Chern-trivial $t=0$ character of the feature precludes
topological order. The data therefore suggest the $(0,1/3)$ minimum is a
moir\'e-commensurate generalized Wigner crystal. A disorder-pinned (glassy) solid would also suppress low-frequency
dissipation, but it has no reason to select these particular moir\'e fillings. What points to a crystal is
the commensurate pattern itself rather than the minimum~(S5).
The $1/3$ crystal also appears through its particle--hole partners $(1,-1/3)$ and $(-1,1/3)$, which fan away
from the $(0,\pm1/3)$ lines across half filling as the field rises. Minima fall on them once they have
separated, on $(-1,1/3)$ at $17.25$ and $18$~T and on $(1,-1/3)$ at $15.4$ and $18$~T, just above half
filling. Their
presence on both carriers is expected for the commensurate crystal, favored over a broad window straddling
half filling~\cite{dasilva2016}.

The weaker minima cannot be so
clearly defined. Each falls near the next commensurate fractions drawn in Fig.~\ref{fig:crystal}, such as
$1/4$ and $2/5$, and could reflect a further crystal commensuration, a general integer Hofstadter gap, or a
mixture the present data cannot separate. The balance tips slightly toward the crystal, since these minima
fall on the commensuration hierarchy~\cite{regan2020,xu2020,jin2021}. The next fractions down, such as
$2/5$, coincide with the data only where their lines cross the dominant $1/3$ lines, and there the stronger
$1/3$ commensuration is the natural assignment, so no weaker fraction is singled out on its own.

The relative prominence of these commensurate fillings is not arbitrary. A moir\'e-registered crystal's commensuration
energy falls with the denominator $q$ of $s=p/q$ and with how poorly its charge order registers with the
triangular lattice. The fillings $s=1/3$ and $2/3$ are singular---their crystals tile the lattice as an
isotropic $\sqrt{3}\times\sqrt{3}$ pattern---and form the most robust generalized Wigner crystals, in both
theory~\cite{mattykim2022} and experiment~\cite{regan2020,xu2020}. Larger denominators cannot tile
isotropically and instead form weaker stripe and dimer orders ($1/2$, $1/4$, $2/5$, $3/5$)~\cite{jin2021,mattykim2022},
while still larger ones ($1/5$, $1/6$, $2/7$) require large supercells stabilized by the long-range Coulomb
interaction~\cite{huang2021,xu2020} and are the most fragile. The line thicknesses in
Fig.~\ref{fig:crystal} encode this expected hierarchy; the observed minima follow it, with $s=1/3$
dominant at every magnetic field and the higher-denominator fractions resolved only as shallow minima.

\textit{Discussion.---} The data suggest the resolved minimum is a moir\'e-commensurate $s=1/3$ crystal, as DaSilva,
Jung, and MacDonald predict~\cite{dasilva2016}, with $s=1/3$ most favorable at the intermediate fillings
where our feature sits. Their predicted onset for the fixed-density trajectories, $\phi/\phi_0\gtrsim0.8$, we reach only at our
highest field, yet the minimum persists across the whole range down to $15.4$~T ($\phi/\phi_0=0.69$),
well below that threshold, consistent with the well-developed moir\'e potential of this closely aligned
device ($A\approx186$~nm$^2$) lowering the onset flux. The crystal is marginal throughout, which is why the feature is a shallow
minimum rather than a hard gap, most prominent near the center of our range. The definitive signature
of a pinned crystal is a sharp pinning-mode resonance. We searched for one and did not observe it, bounding
any resonant excess below $\approx25$~mdB across $0.5$--$9$~GHz (S4). If a pinning mode exists, it lies
outside our band, below our resolution, or damped below detection at our drive power.

\textit{Summary.---} Microwave dissipation in the lowest Landau level is a disorder-broadened dome
whose width is a field-independent density inhomogeneity $\delta n$, smaller than the zero-field
transport value ($\delta n\lesssim\delta n_{\rm tr}$). Riding on it, we observe a family of
minima that hold a fixed per-cell density (most prominent a few percent below half
filling and drifting toward charge neutrality with field), recurring on both carriers at every magnetic
field and matching no fractional QHE or integer Hofstadter origin. The one origin that tracks them across
all magnetic fields is a moir\'e-commensurate $(t,s)=(0,1/3)$ electron crystal of one electron per three
moir\'e cells~\cite{dasilva2016}. Additional weaker minima are consistent with other fractional moir\'e
fillings, though we cannot exclude an interplay with integer Hofstadter gaps. No pinning-mode
resonance is observed within our band.

\begin{acknowledgments}
This material is based upon work supported by the National Science Foundation under Grant No.
DMR-1943389. K.W. and T.T. acknowledge support from the JSPS KAKENHI (Grant Numbers 21H05233 and
23H02052) and the World Premier International Research Center Initiative (WPI), MEXT, Japan. Part of
the work was carried out at the Singh Center for Nanotechnology, which is supported by the NSF
National Nanotechnology Coordinated Infrastructure Program under Grant No. NNCI-2025608. A portion of
this work was performed at the National High Magnetic Field Laboratory, which is supported by National
Science Foundation Cooperative Agreement No. DMR-2128556 and the State of Florida.
\end{acknowledgments}

\end{document}


\title{Supplemental Material:\\ Microwave spectroscopic evidence of a
moir\'e-commensurate electron crystal in monolayer graphene}
\author{A.~Solodovnyk \textit{et al.}}
\noaffiliation
\date{\today}
\maketitle

\section{Device, fabrication, and calibration}
\label{sm:device}

The device is a monolayer graphene flake encapsulated in hexagonal boron nitride, gated by a graphite
back gate, and placed in the gap of a Au coplanar waveguide. Transmission $S_{21}(f)$ is measured with
a vector network analyzer over $f=0.5$--$9$~GHz at $\sim50$~mK in linear response.

The gate lever arm is measured, not assumed: $\alpha=1.536\times10^{11}$~cm$^{-2}$/V, obtained from the
charge-neutrality-free $\nu=-1$ to $+1$ span of $5.668$~V at $18$~T --- a span that requires neither the
charge-neutrality point nor a registration offset --- as $\alpha=2(eB/h)/\Delta V_g$. With the
AFM-measured gate-hBN thickness of $120$~nm this lever corresponds to a dielectric constant
$\epsilon_{\rm hBN}=3.33$, a consequence rather than an input. From $\alpha$ and the $B=0$ transport
secondary-Dirac span the moir\'e unit-cell area is $A=186.3$~nm$^2$, the period $\lambda=14.67$~nm, the
full-band density $n_s=2.147\times10^{12}$~cm$^{-2}$, and the flux per cell $\phi/\phi_0=0.04505\,B$,
placing half flux ($\phi/\phi_0=1/2$) at $11.10$~T, in agreement with the observed pinch of the filling
lines in the Wannier diagram. The microwave drive was held in the linear-response regime, verified by the
power-independence of $D(\nu)$ and of the minima positions.

\section{Dome fit and the disorder broadening}
\label{sm:dome}

The main text reports the dome width phenomenologically, as a broadened-level width $\eta$ converted to
a density $\delta n=\eta\,n_\phi$ ($n_\phi=eB/h$). Here we give the line-shape model behind that width
and the reason we do not lean on it in the main text.

\paragraph*{Model.} We fit each side of $D(\nu)$ with a semi-elliptic density of states, the
self-consistent Born approximation (SCBA) form of half-width $W_{\rm h}$, convolved with a Gaussian
of width $\eta$, plus a linear background; the electron and hole sides are fit independently, so no
electron--hole symmetry is imposed. The Gaussian width is locked at $\eta=0.03$ and the centers are
free; $W_{\rm h}$ and the amplitudes are fit. The fitted half-width is field- and side-independent,
$W_{\rm h}\approx0.41$--$0.42$ on every clean two-sided scan from $15.4$ to $18$~T.

\paragraph*{Why $\eta$ is locked.} The Gaussian width $\eta$ and the semi-ellipse half-width
$W_{\rm h}$ are degenerate in the line shape --- fitting more of the shoulders lowers $\eta$ and
widens $W_{\rm h}$ with the same final curve --- so a free per-side $\eta$ is unidentifiable and
unstable, flipping sides between fields. Locking $\eta$ breaks the degeneracy and lets $W_{\rm h}$
carry the width; the electron/hole disorder asymmetry is instead read from the transport mobility
ratio $\mu_e/\mu_h=1.87$ and the dome shape.

\paragraph*{Disorder density.} Writing $\eta=\delta n/n_\phi$, the field-independent quantity is
$\delta n$:

\begin{table}[htb]
\caption{\label{tab:dn}Disorder density from the per-field dome fits.}
\begin{ruledtabular}
\begin{tabular}{cccc}
$B$ (T) & $\eta$ & $n_\phi$ (cm$^{-2}$) & $\delta n$ (cm$^{-2}$) \\
\colrule
$16.4$ & $0.033$ & $3.97\times10^{11}$ & $1.31\times10^{10}$ \\
$18.0$ & $0.030$ & $4.35\times10^{11}$ & $1.31\times10^{10}$ \\
\end{tabular}
\end{ruledtabular}
\end{table}

\noindent Both fields return $\delta n=1.31\times10^{10}$~cm$^{-2}$; $\eta$ carries only the mild $1/B$
trend.

\begin{figure}[htb]
\includegraphics[width=\linewidth]{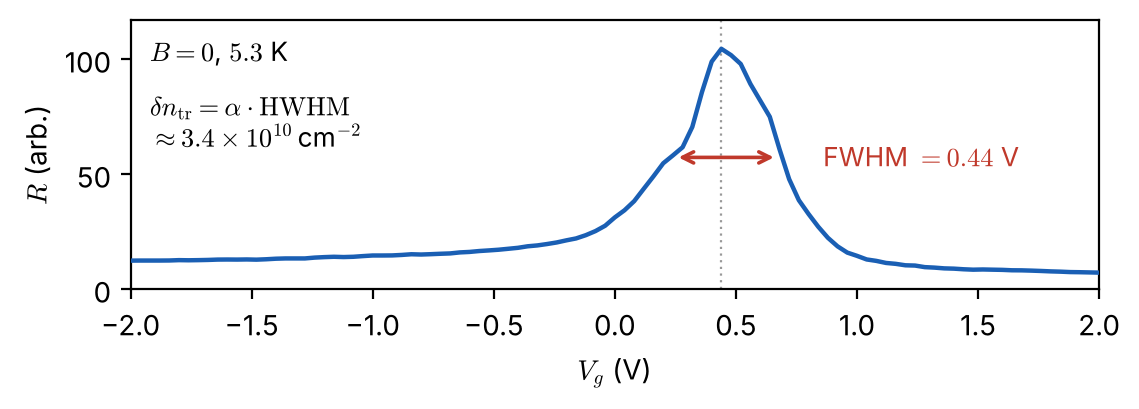}
\caption{\label{fig:dntr}\textbf{Zero-field disorder density.} Four-probe transport at $B=0$,
$5.3$~K (lock-in through a $10$~M$\Omega$ series load; signal $\propto R$) versus gate voltage
$V_g$. The charge-neutrality peak has FWHM $=0.44$~V; its half-width, scaled by the lever arm
$\alpha=1.536\times10^{11}$~cm$^{-2}$/V, gives the static charge-inhomogeneity density
$\delta n_{\rm tr}\approx3.4\times10^{10}$~cm$^{-2}$.}
\end{figure}

\paragraph*{Zero-field transport width.} The independent disorder scale comes from four-probe
transport at $B=0$ and $5.3$~K, measured with a lock-in through a $10$~M$\Omega$ series load so the
recorded signal is proportional to the sample resistance $R$ (Fig.~\ref{fig:dntr}). The
charge-neutrality peak has a half-width at half maximum of $0.22$~V; converted through the measured
lever arm $\alpha=1.536\times10^{11}$~cm$^{-2}$/V this gives the static charge-inhomogeneity density
$\delta n_{\rm tr}=\alpha\cdot\mathrm{HWHM}\approx3.4\times10^{10}$~cm$^{-2}$. The microwave dome
therefore reports about $0.4$ of the transport disorder. The two are the same charge inhomogeneity
seen in different regimes: the $B=0$ width is a DC, spatially averaged measure of the static
landscape, whereas the high-field $\delta n$ is dominated by the delocalized states near the level
center, so the microwave value is expected to be the smaller of the two.

\paragraph*{Validity.} The SCBA semi-ellipse follows from short-range ($\delta$-correlated) disorder, a
single isolated Landau level ($\hbar\omega_c\gg$ the broadening), and weak (Born) scattering in a
Gaussian random potential --- conditions established for the GaAs two-dimensional electron gas that may
not transfer to moir\'e-aligned graphene. For that reason the main text claims only the phenomenological
width and its tie to the $B=0$ inhomogeneity, and does not rest on the SCBA line shape.

\section{Identification of minima, field dependence, and commensuration test}
\label{sm:dips}

The minimum depth is measured on unsmoothed $D(\nu)$ and does not depend on the dome fit; the
position is located from the full-width fit only, to avoid pre-assigning fractions. The resolved
electron-side minimum is present at every field, moving from $\nu=+0.468$ at $16.4$~T toward charge
neutrality at higher field. Its Landau filling drifts downward with field along a line of constant per-cell
density, the signature of a $t=0$ commensurate state, rather than sitting at a fixed Landau filling as a
fractional quantum Hall or Jain state would.

Table~\ref{tab:dips} lists the local minima located across the four clean two-sided scans,
with each minimum's Landau filling $\nu$, span (full width used for the error bars in Fig.~3 of the main
text), per-cell filling $\nu_P=|\nu|\,(\phi/\phi_0)$, nearest low-order commensurate fraction $s$, and
deviation $|\nu_P-s|$. Several minima sit close to $\nu_P=1/3$, including the electron-side feature that
recurs at every field; the remaining minima fall near other low-order fractions and are weaker and less
regular.

\begin{table}[htb]
\caption{\label{tab:dips}Local minima of $D(\nu)$ across the four two-sided scans. $\nu_P=|\nu|\,
(\phi/\phi_0)$ is the per-cell filling; $s$ is the nearest low-order commensurate fraction and dev
$=|\nu_P-s|$.}
\begin{ruledtabular}
\begin{tabular}{lcccccc}
scan & $B$ (T) & $\nu$ & span & $\phi/\phi_0$ & $\nu_P$ & $s$ (dev) \\
\colrule
050 & 15.4 & $-0.493$ & 0.049 & 0.694 & 0.342 & $1/3$ (0.009) \\
050 & 15.4 & $-0.369$ & 0.056 & 0.694 & 0.256 & $1/4$ (0.006) \\
050 & 15.4 & $-0.264$ & 0.045 & 0.694 & 0.183 & $1/4$ (0.067) \\
050 & 15.4 & $+0.290$ & 0.045 & 0.694 & 0.201 & $1/4$ (0.049) \\
050 & 15.4 & $+0.404$ & 0.045 & 0.694 & 0.280 & $1/4$ (0.030) \\
050 & 15.4 & $+0.468$ & 0.045 & 0.694 & 0.325 & $1/3$ (0.008) \\
050 & 15.4 & $+0.550$ & 0.050 & 0.694 & 0.382 & $2/5$ (0.018) \\
050 & 15.4 & $+0.668$ & 0.059 & 0.694 & 0.463 & $1/2$ (0.037) \\
050 & 15.4 & $+0.785$ & 0.051 & 0.694 & 0.545 & $1/2$ (0.045) \\
049 & 16.4 & $-0.808$ & 0.045 & 0.739 & 0.597 & $3/5$ (0.003) \\
049 & 16.4 & $-0.711$ & 0.054 & 0.739 & 0.525 & $1/2$ (0.025) \\
049 & 16.4 & $-0.447$ & 0.079 & 0.739 & 0.330 & $1/3$ (0.003) \\
049 & 16.4 & $-0.312$ & 0.066 & 0.739 & 0.230 & $1/4$ (0.020) \\
049 & 16.4 & $-0.205$ & 0.063 & 0.739 & 0.151 & $1/4$ (0.099) \\
049 & 16.4 & $+0.233$ & 0.045 & 0.739 & 0.172 & $1/4$ (0.078) \\
049 & 16.4 & $+0.310$ & 0.082 & 0.739 & 0.229 & $1/4$ (0.021) \\
049 & 16.4 & $+0.468$ & 0.045 & 0.739 & 0.346 & $1/3$ (0.013) \\
049 & 16.4 & $+0.599$ & 0.050 & 0.739 & 0.442 & $2/5$ (0.042) \\
049 & 16.4 & $+0.652$ & 0.090 & 0.739 & 0.482 & $1/2$ (0.018) \\
046 & 17.25 & $-0.808$ & 0.062 & 0.777 & 0.628 & $3/5$ (0.028) \\
046 & 17.25 & $-0.684$ & 0.063 & 0.777 & 0.531 & $1/2$ (0.031) \\
046 & 17.25 & $-0.557$ & 0.059 & 0.777 & 0.433 & $2/5$ (0.033) \\
046 & 17.25 & $-0.404$ & 0.045 & 0.777 & 0.314 & $1/3$ (0.019) \\
046 & 17.25 & $-0.217$ & 0.102 & 0.777 & 0.169 & $1/4$ (0.081) \\
018 & 18.0 & $-0.825$ & 0.045 & 0.811 & 0.669 & $2/3$ (0.002) \\
018 & 18.0 & $-0.757$ & 0.059 & 0.811 & 0.614 & $3/5$ (0.014) \\
018 & 18.0 & $-0.590$ & 0.045 & 0.811 & 0.478 & $1/2$ (0.022) \\
018 & 18.0 & $-0.260$ & 0.054 & 0.811 & 0.211 & $1/4$ (0.039) \\
018 & 18.0 & $+0.264$ & 0.047 & 0.811 & 0.214 & $1/4$ (0.036) \\
018 & 18.0 & $+0.438$ & 0.049 & 0.811 & 0.355 & $1/3$ (0.022) \\
018 & 18.0 & $+0.569$ & 0.073 & 0.811 & 0.461 & $1/2$ (0.039) \\
018 & 18.0 & $+0.754$ & 0.077 & 0.811 & 0.611 & $3/5$ (0.011) \\
018 & 18.0 & $+0.876$ & 0.064 & 0.811 & 0.711 & $2/3$ (0.044) \\
\end{tabular}
\end{ruledtabular}
\end{table}

\paragraph*{Commensuration.} The recurring electron-side minimum, and its charge-conjugate partner on
the hole side, follow the fixed-density $s=1/3$ lines $(0,\pm1/3)$: they hold a constant per-cell filling
$\nu_P=1/3$, drift as $1/B$, and appear at every field. This fixed-density tracking, not the depth of any
single minimum, is what marks the commensuration, and $s=1/3$ is the commensuration DaSilva \textit{et al.}
identify \textit{a priori} as most favorable at these intermediate fillings. We present it as the
organizing principle for the minima rather than a statistical detection; the remaining features fall
near other low-order fractions and are weaker and less regular.

\section{Pinning-mode search and the filling constraint}
\label{sm:pinning}

At the filling of the minimum, the frequency dependence is statistically indistinguishable from
that of the flanking dome, bounding any resonant excess below $\approx25$~mdB across $0.5$--$9$~GHz; the
minimum is thus an amplitude modulation of the dome, not a resonance, and any pinning mode lies
outside our band or below our resolution.

A general filling constraint favors a solid: a gapped state at a fractional filling per moir\'e cell
cannot be a simple, featureless insulator, but must either break translational symmetry (a crystal) or
carry intrinsic topological order (a Lieb--Schultz--Mattis-type argument). A nonzero Hall response would
also satisfy the constraint, but the feature sits on the Hall-free $t=0$ line, closing that route.

\section{Crystal versus disorder-pinned glass}
\label{sm:glass}

Any pinned charge solid suppresses low-frequency dissipation, because pinned charge does not dissipate
until it is depinned. A clean moir\'e crystal and a disorder-pinned (glassy) solid therefore produce
the same kind of minimum in $D(\nu)$, and that minimum \emph{by itself} does not distinguish them; we
concede this. What distinguishes them is the \emph{pattern}, not the minimum. The dominant feature
sits at and tracks a fixed commensurate per-cell density $\nu_P=1/3$ (the isotropic
$\sqrt{3}\times\sqrt{3}$ triangular crystal) at every field, and the weaker features follow the
triangular commensuration hierarchy. A generic disorder glass has no reason to select those particular
moir\'e fillings or to reproduce that hierarchy; disorder-pinned solids in the GaAs two-dimensional
electron gas, moreover, pin preferentially at the dilute edges of the level rather than near half
filling, which is where our feature lies.

The measured energy scales place the state more precisely. At $16.4$~T the Coulomb scale is
$e^2/4\pi\epsilon\ell_B\approx45$~meV ($\epsilon_r\approx5.0$, the geometric mean of the anisotropic
hBN constants; $\ell_B\approx6.3$~nm), while the disorder amplitude obtained by converting the density
widths through the graphene Dirac density of states, $E=\hbar v_F\sqrt{\pi\,\delta n}$, is
$\approx13$~meV from the microwave dome and $\approx20$~meV from the $B=0$ transport width. Interactions
thus exceed disorder by only $\approx2$--$3\times$: the electrons form a solid rather than a
disorder-smeared liquid, but not by a wide margin, and the disorder is comparable to the smaller
moir\'e lock-in (commensuration) energy. The state is accordingly best described not as a defect-free
crystal nor as a featureless glass, but as a marginal, commensurate-correlated pinned solid --- the
moir\'e biasing \emph{where} it pins and disorder setting \emph{how well} --- consistent with the main
text calling the crystal marginal throughout. In this sense the glass is a limit of the
disorder-pinned Wigner solid already under discussion rather than a separate rival to exclude. (The
triangular lattice constant implied by $\nu_P=1/3$ is $a=\sqrt{3}\,\lambda=25.4$~nm, as required for a
$\sqrt{3}\times\sqrt{3}$ superstructure of the $14.7$~nm moir\'e; these estimates are
order-of-magnitude.)

Two features of the present data lean commensurate rather than glassy: the in-band resonant excess is
bounded below $25$~mdB (S4), tighter than a weakly disorder-pinned glass would likely show, and the
minimum is filling-selective rather than smooth in $\nu$, following the moir\'e commensuration rather
than a disorder-set scale.

\section{Relaxation frequency $f^\ast$ and the composite-fermion scale}
\label{sm:fstar}

\begin{figure}[htb]
\includegraphics[width=\linewidth]{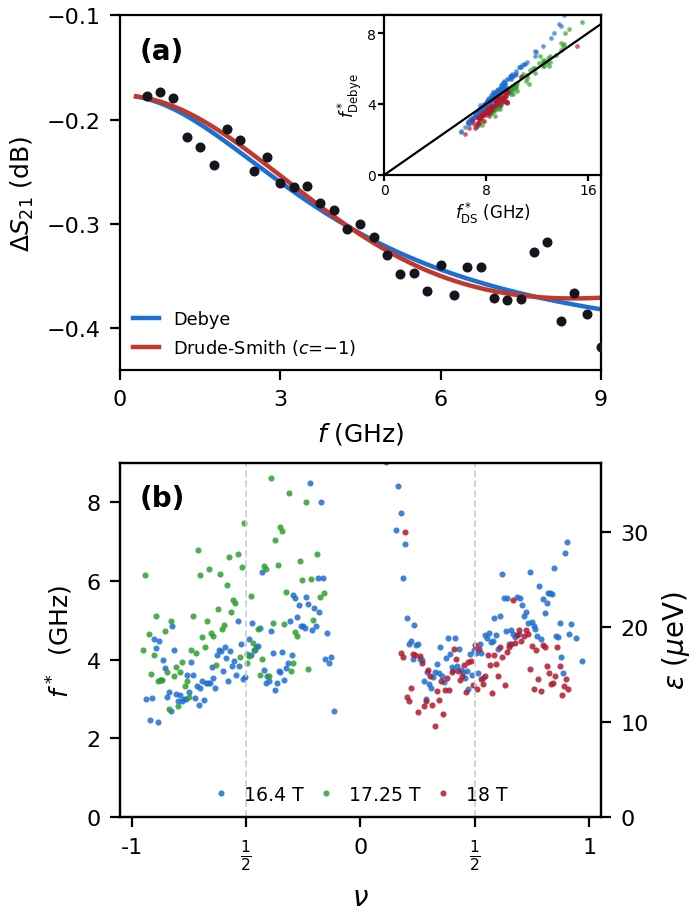}
\caption{\label{fig:sfstar}\textbf{Relaxation frequency.} (a)~The $\nu=1/2$ dissipative spectrum
$\Delta P(f)$ (points) with Debye (blue) and Drude--Smith ($c=-1$, red) fits; inset, the Debye
rollover frequency versus the Drude--Smith value, $f^\ast_{\rm Debye}=(0.50\pm0.03)\,f^\ast_{\rm DS}$
(line, slope $1/2$). (b)~$f^\ast(\nu)$ across the lowest Landau level at all three fields, flat and
featureless at $f^\ast\approx4.3$~GHz ($\sim18$~$\mu$eV), with no structure at half filling and no
fractional imprint. This material was formerly a panel of the main-text dome figure.}
\end{figure}

The characteristic rollover frequency $f^\ast$ is extracted at each filling from the frequency
dependence of $\Delta P(f)$ (the transmitted power in dB, as in the main text) by Debye and Drude--Smith fits (the Debye value is
$0.50\pm0.03$ times the Drude--Smith value). Across the lowest Landau level $f^\ast$ is flat and
featureless, $f^\ast\approx4.3\pm0.5$~GHz ($\sim18$~$\mu$eV), with no structure at half filling and no
fractional imprint at any field, and it coincides at $16.4$ and $18$~T (Fig.~\ref{fig:sfstar}). A flat
$f^\ast$ is what a disorder-broadened dome produces: it reflects the single-particle relaxation of the
broadened level, not an interaction-set scale that would track filling. Disordered and pinned
electron systems carry their own filling- and disorder-dependent characteristic frequency
[L.~W.~Engel \textit{et al.}, Phys.~Rev.~Lett.~\textbf{71}, 2638 (1993); H.~A.~Fertig,
Phys.~Rev.~B~\textbf{59}, 2120 (1999); M.~M.~Fogler and D.~A.~Huse, Phys.~Rev.~B~\textbf{62}, 7553
(2000)], so a flat, featureless $f^\ast$ is \emph{consistent with} disorder broadening rather than
proof against an interaction-set scale; we therefore read the absence of any filling-tracking scale
as supporting, not proving, the disorder picture.

\paragraph*{The composite-fermion frequency scale (Kohn/CF-CR).} By Kohn's theorem the long-wavelength
cyclotron resonance sits at the bare cyclotron frequency $\omega_c=eB/m$ and is unrenormalized by
electron--electron interactions, so it carries no information about the correlated state. The
interaction-relevant analog for a composite-fermion liquid is the composite-fermion cyclotron
resonance (CF-CR) at the effective field $B^\ast=B-B_{1/2}$, with a frequency
$f_{\rm CF}=eB^\ast/2\pi m^\ast_{\rm CF}$ set by the composite-fermion mass $m^\ast_{\rm CF}$. This is
the scale a composite-fermion Fermi liquid would imprint on $f^\ast$, and its absence (a flat $f^\ast$)
is the statement in the main text. A numerical estimate of $f_{\rm CF}$ from the effective field and a
Coulomb-scale composite-fermion mass is kept with the reserve analyses and is not reproduced here.

\section{Diophantine classification of the candidate states}
\label{sm:classify}

The candidate states are organized by their Diophantine indices $(t,s)$ of Eq.~(1), with the
band-filling (Bloch) index $s$ as the primary label and the Hall (Chern) index $t$ as a secondary one.
A state with $s=0$ is a pure Landau-level quantum Hall state with no moir\'e commensuration, integer-%
quantized when $t$ is an integer and fractional (Laughlin, Jain, or the $\nu=1/2$ composite-fermion sea)
when $t$ is fractional. A state with integer $s\neq0$ is a single-particle Hofstadter gap of the fractal
spectrum. A moir\'e-commensurate crystal has a fractional $s=p/q$, the number of crystallizing carriers
per cell. The crystal is itself Chern-trivial, so its Hall index $t$ carries no intrinsic information: it
merely counts the filled Landau level the carriers are reckoned from, $t=0$ from charge neutrality and
$t=\pm1$ from the next filled level, which places the electron crystal and its particle--hole conjugate
hole crystal on the same footing (Table~\ref{tab:classify}).

\begin{table}[htb]
\caption{\label{tab:classify}Diophantine classification of the candidate states by indices $(t,s)$ of
Eq.~(1), with $\sigma_{xy}=t\,e^2/h$. The band-filling index $s$ is the primary label; a nonzero $s$
means commensuration with the moir\'e superlattice. Within a given $s$, an integer $t$ marks an
incompressible/insulating state and a fractional $t$ a fractional quantum Hall liquid.}
\begin{ruledtabular}
\begin{tabular}{lll}
$(t,s)$ & $\sigma_{xy}$ & state \\
\colrule
$t$ int.\ or frac., $s=0$ & $t\,e^2/h$ & quantum Hall (IQHE / FQHE) \\
$t$ int., $s$ int.\ $\neq0$ & integer & Hofstadter (fractal) gap \\
$t$ int.\ ($=$ filled LL), $s=p/q$ & integer & moir\'e-commensurate crystal \\
$t$ frac., $s=p/q$ & fractional & fractal (replicated) FQHE \\
\end{tabular}
\end{ruledtabular}
\end{table}

In our field range the $s=0$ row holds the $\nu=\pm1$ gaps $(\pm1,0)$ and the $\nu=1/2$
composite-fermion sea $(1/2,0)$; the integer-$s$ row holds the tested first-order Hofstadter gaps
$(-1,+1)$, $(-2,+1)$, and $(-3,+2)$; and the crystal row holds $(0,\pm1/3)$ and $(0,\pm2/3)$, counted
from charge neutrality, together with their particle--hole conjugates $(1,-1/3)$ and $(-1,+1/3)$, counted
from the filled level. Two one-line tests follow from the table: a crystal is set apart from a Hofstadter
gap by a fractional versus integer $s$, and from a fractional quantum Hall liquid---including the fractal
FQHE---by an integer versus fractional $t$. The indices are kinematic and do not distinguish a crystal
from a hypothetical correlated state at the same $(t,s)$; that distinction rests on the dissipative
signature and the pinning mode (S4, S5).